\documentclass[aps,prl,twocolumn,showpacs,footinbib]{revtex4}
\usepackage{amssymb,amsbsy,graphicx,times,subfigure,marvosym,color,hyperref}
\usepackage[latin1]{inputenc}
\begin{document}
\draft

\title{Parity-dependent proximity effect\\ in superconductor/antiferromagnet heterostructures}
\author{J. W. A. Robinson, G\'{a}bor Hal\'{a}sz, and M. G. Blamire}
\address{Department of Material Science, University of Cambridge, Pembroke Street, Cambridge CB2 3QZ, UK}

\begin{abstract}
We report the effect on the superconducting transition temperature
($T_c$) of a Nb film proximity coupled to the synthetic
antiferromagnet Fe/$\{$Cr/Fe$\}_{N-1}$. We find that there is a
parity dependance of $T_c$ on the total number of Fe layers, $N$;
locally $T_c$ is always a maximum when $N$ is even, and a minimum
when $N$ is odd. The Fe electron mean free path and coherence length
are indicative of dirty limit behavior; as such, we numerically
model our data using the linearized Usadel equations with good
correlation.

\end{abstract}

\date{\today}

\maketitle

During the past fifteen years the coupling of thin film
superconductors (S) to ferromagnets (FM) \cite{Review} has been
extensively studied. In the most simplistic sense, these electron
coupling phenomena can be considered to be mutually exclusive with
spin alignment enforced by ferromagnetic exchange and anti-parallel
spin alignment necessary for singlet BCS superconductivity.
According to theory, the penetration depth of the superconducting
proximity effect (SPE) into normal metals (NM) is governed by the
electron phase breaking length. When NM is substituted with a FM,
the singlet superconducting correlations and phase coherent effects
are destroyed by the exchange interaction $I$ within the FM
coherence length $\xi_{f}\propto \sqrt{1/I}$ because of the
splitting of spin-up and spin-down conduction bands. For S/FM
bilayers, a consequence of $I$ on S is the non-monotonic dependence
of $T_c$ on FM layer thickness ($d_{f}$) for when the superconductor
is thinner than its BCS coherence length \cite{Jiang1995}. This can
be understood in terms of an interference effect between the
transmitted singlet pair wave function through the S/FM interface
with the wave reflected at the opposite FM interface.

Recently, Andersen \emph{et al.} \cite{Andersen} predicted a SPE
dependence on the number of antiferromagnetic (AF) atomic planes
($N$) in S/AF/S junctions such that the junction's ground state is
$0$ or $\pi$ depending on whether $N$ is even or odd.
Experimentally, this SPE dependence on $N$ is difficult to realize
because it demands atomic thickness control of the AF. So far,
experiments looking at the proximity of S to AFs have focused on
thicker AF layers \cite{SAFS} where this parity dependence does not
exist. A way to investigate a parity dependent SPE is to use
synthetic antiferromagnets (SAFs) which exploit the AF coupling of
FM layers separated by a NM spacer \cite{ParkinPRL6423041990}.
\begin{figure}[b!]
\centering
\includegraphics[width=6cm]{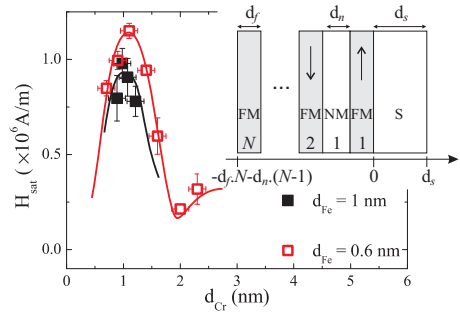}
\caption{(color online). Dependence of $H_{sat}$ on $d_{Cr}$ in
Nb(20nm)/Fe/$\{$Cr/Fe$\}_{\times 7}$ films for: $d_{Fe}$=0.6 nm
(\textcolor[rgb]{1.00,0.00,0.00}{$\square$}); and $d_{Fe}$=1.0 nm
($\blacksquare$). Solid lines are to guide the eye. Inset:
illustration of S/FM/$\{$NM/FM$\}_{N-1}$ structure with $N$ FM
layers and $N-1$ NM layers. Vertical arrows indicate FM polarization
direction.\label{Hsatvsdcr}}
\end{figure}

So far, the proximity effect of SAFs coupled to S materials have not
been considered although the $T_c$ dependence of complicated
multilayers have been extensively studied \cite{FeMulti}; for
example, the control of $T_c$ in pseudo-spin valve FM/S/FM
structures has been proposed \cite{PSVTHEORY} and realized
experimentally \cite{PSV} with mK differences in $T_c$ between
parallel (P) and anti-parallel (AP) FM configurations. The most
similar structure to the one we study here was proposed by Oh
\emph{et al.} \cite{Sangjun} in which the S layer is in proximity to
decoupled FM/NM/FM. In this Letter, we report the
proximity-suppression of $T_c$ of a Nb film coupled to a
Fe/$\{$Cr/Fe$\}_{N-1}$ SAF. In doing this, we find that $T_c$ has a
pronounced parity dependence on $N$ such that when $N$ is even,
$T_c$ is a local maximum and, conversely, when $N$ is odd, $T_c$ is
a local minimum.


Films were grown in Ar (1.5 Pa) on oxidized (120 nm) Si (100)
(surface area: 5-10$\times5$ mm$^2$) in a diffusion-pumped ultrahigh
vacuum sputter deposition system, consisting of the following: three
dc magnetrons; a computer operated (rotating) sample table; a liquid
N$_2$ cooling jacket; and a residual gas analyzer. The vacuum system
was baked-out overnight prior to each experiment, reaching a base
pressure of 1-4$\times$10$^{-6}$ Pa. One hour before depositing, the
system was cooled with liquid N$_2$ giving a final base pressure of
1-3$\times$10$^{-8}$ Pa, a residual $O_2$ pressure of
$\leq$4$\times$10$^{-9}$ Pa, and an outgassing rate of $\approx$
1$\times$10$^{-8}$ Pa s$^{-1}$. Targets (Nb, Fe, and Cr of 99.9$\%$
purity), were pre-sputtered for $\sim$ 15 minutes to remove
contaminants from their surfaces and to further reduce the base
pressure of the vacuum chamber by getter sputtering. To control film
thickness ($d$) and to ensure clean interfaces, films were grown in
a single sweep by rotating the substrates around the symmetry of the
chamber under stationary targets. Growth rates were pre-calibrated
by growing films on patterned substrates and, from a lift-off
step-edge, $d$ for each material was established using an atomic
force microscope. Typical growth rates with a sweep speed of 1 rpm
per pass were: 1.0 nm for Cr; 0.6 nm for Fe; and 1.4 nm for Nb. The
average film roughness was $\leq$ 3 {\AA} over 1 $\mu$m.

\begin{figure}[b]
\centering
\includegraphics[width=6cm]{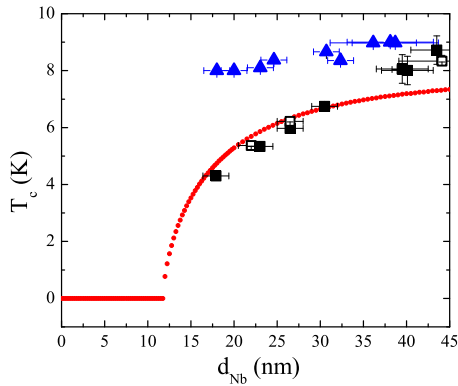}
\caption{(color online). Dependence of $T_c$ on $d_{Nb}$: bare Nb
($\textcolor[rgb]{0.00,0.00,1.00}{\blacktriangle}$); two
Nb/Fe/$\{$Cr/Fe$\}_{\times 7}$ sets for $d_{Fe}=$ 0.6 nm and
$d_{Cr}=$ 0.9 nm ($\square$ and $\blacksquare$); theory fit
($\textcolor[rgb]{1.00,0.00,0.00}{\bullet}$) explained in text.
\label{TcCalibration}}
\end{figure}

The antiferromagnetic interlayer exchange coupling (AFC) between the
Fe layers was investigated in two sets of Nb/Fe/$\{$Cr/Fe$\}_{\times
7}$ films: for $d_{Fe}$=0.6 nm; and for $d_{Fe}$=1.0 nm. $d_{Nb}$
was constant (20$\pm1$ nm) while $d_{Cr}$ varied in the 0.5-2.5 nm
range. A 2.8 nm capping layer of Nb was grown on the top Fe layer.
To quantify AFC in these films and to optimize $d_{Cr}$ to give the
largest AFC energy $J$ (given by $-4J=\mu_0 H_{sat} M d_{Fe}$ where
$M$ is the magnetization of the Fe), the required saturating field
$H_{sat}$ needed to align the Fe layers was measured with a
vibrating sample magnetometer at room temperature as a function of
$d_{Cr}$; see Fig.\ref{Hsatvsdcr}. For $d_{Fe}=$ 0.6 nm, $H_{sat}$
is a maximum value of (1.16$\pm$0.04)$\times10^6$ A/m for $d_{Cr}=$
1.0$\pm0.2$ nm with $M \approx$ 0.8$\times$10$^6$ A/m (similar to
\cite{Robinson2007}). This corresponds to $-J\approx$
1.8$\times$10$^{-4}$ Jm$^{-2}$, which compares well to previously
reported energies \cite{ParkinPRL6423041990}. To achieve the most
efficient AFC, we chose $d_{Fe}=$ 0.6 nm and $d_{Cr}=$ 0.9 nm for
the appropriate layer thicknesses.

To establish a $d_{Nb}$ range that is strongly affected by the
presence of the SAF film, the $T_c$ of both bare Nb and
Nb/Fe/$\{$Cr/Fe$\}_{\times7}$ films was measured in the 15-45 nm
range; see Fig. \ref{TcCalibration}. The critical $d_{Nb}$ is
$\approx$ 12 nm. $T_c$ was estimated by measuring the resistance of
a film as a function of temperature ($R(T)$) using a standard four
contact technique; for this, two instruments were used: a custom
made liquid He dip probe and a pumped He-4 temperature insert. To
make electrical contacts, films were ultrasonically wire-bonded onto
copper carriers with Al wire (25 $\mu$m diameter). Samples were
fixed to their carriers with silver conducting paint. An ac current
of $\pm$10$\mu$A was applied. $T_c$ was estimated from warming curve
data and defined as the mid-point of the $R(T)$ transition.

\begin{figure}[t!]
\centering
\includegraphics[width=6cm]{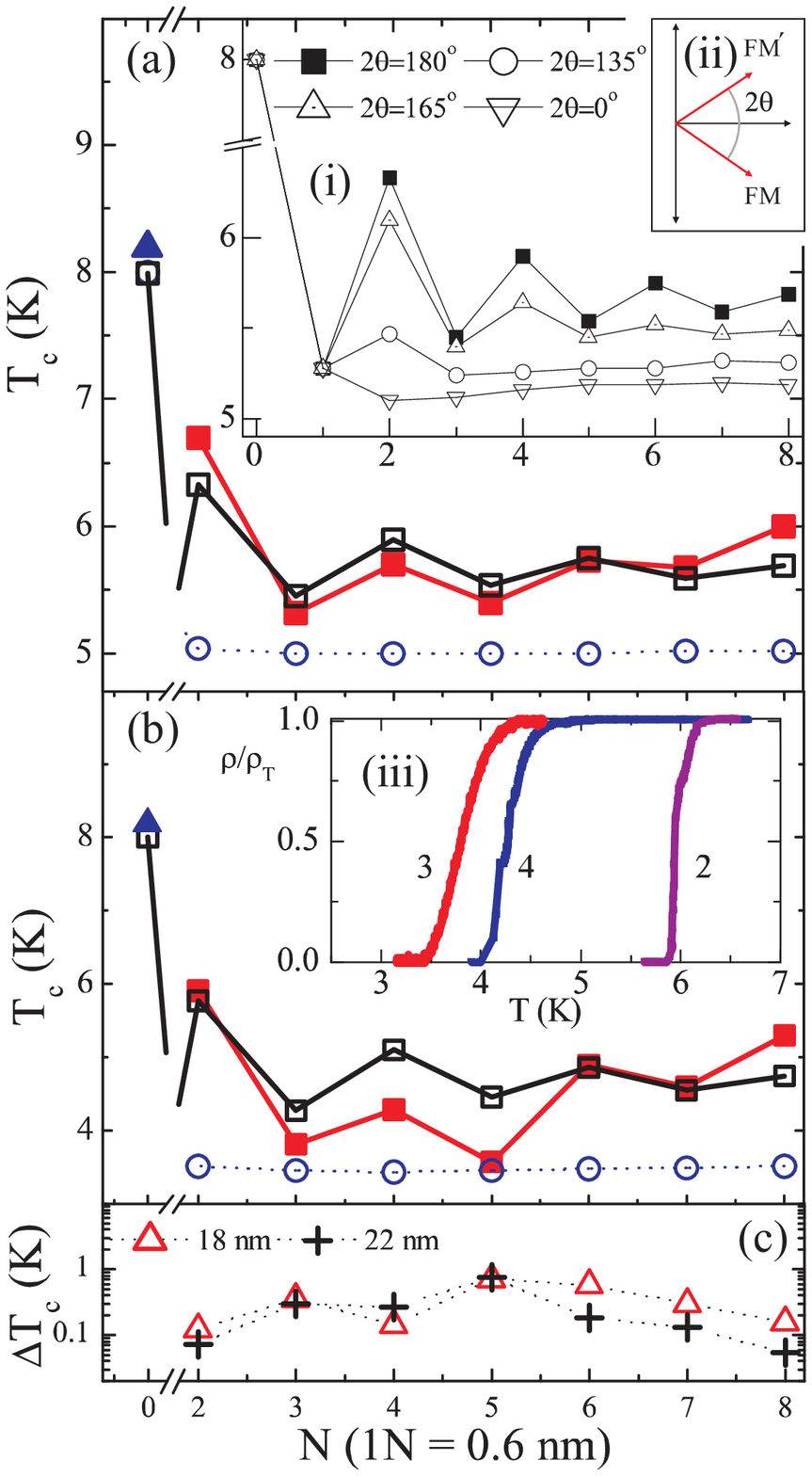}
\caption{(color online). Experimental
($\textcolor[rgb]{1.00,0.00,0.00}{\blacksquare}$) and theoretical
($\square$) dependence of $T_c$ on $N$: (a) $d_{Nb}=$ 22 nm; (b)
$d_{Nb}=$ 18 nm. For comparison, data points
($\textcolor[rgb]{0.00,0.00,1.00}{\odot}$) are for a theoretical
S/FM bilayer and ($\textcolor[rgb]{0.00,0.00,1.00}{\blacktriangle}$)
for bare Nb taken from Fig. 2. Inset: (i) the theoretical
non-collinear case for various $2\theta$ angles between FM layers;
(ii) illustration defining $2\theta$ between two FM layers; (iii)
typical resistivity curves for $N=$ 2,3 and 4 in the function of
$T$. (c) Spread in $T_c$ ($\Delta T_c$) defined as 20$\%$-80$\%$ of
the transition. \label{TcvN}}
\end{figure}

Fig. \ref{TcvN} shows the dependence of $T_c$ on $N$ for two Nb
thicknesses: (a) 22 nm and (b) 18 nm. We see that $T_c$ is a local
maximum when $N$ is even and vice-versa. The magnitude of
$|T_c(N+1)-T_c(N)|$ decreases as $N$ increases: the largest for
$T_c(2)-T_c(3)=$1.38 K in (a) and 2.06 K in (b). Assuming dirty
limit behaviour in the multilayers ($\xi>\ell$), we have adapted the
Usadel equations, as discussed by Fominov \emph{et al.}
\cite{FominovPRB660145072002} and Oh \emph{et al.} \cite{Sangjun},
to model the situation; as illustrated in the inset of Fig.
\ref{Hsatvsdcr}.

The normal Green function is $G = \mathrm{sgn}(\omega_l)$ when
$T\approx T_c$ and the linearized Usadel equations for the anomalous
Green function $F$ are:
\begin{equation}
\frac{\hbar D_s}{2}\frac{d^2 F_s}{dx^2}-\hbar |\omega_l| F_s +
\Delta = 0 \label{usadels}
\end{equation}
inside the S layer,
\begin{equation}
\frac{\hbar D_f}{2}\frac{d^2 F_{fk}}{dx^2}-\hbar |\omega_l| F_{fk} -
i (-1)^k I \mathrm{sgn}(\omega_l)F_{fk}= 0 \label{usadelf}
\end{equation}
inside the $k$th ($k = 1,2,... N$) FM layer and
\begin{equation}
\frac{\hbar D_n}{2}\frac{d^2 F_{nk}}{dx^2}-\hbar |\omega_l| F_{nk} =
0 \label{usadeln}
\end{equation}
inside the $k$th ($k = 1,2,... N-1$) NM layer.

The appropriate diffusivities are $D_s$, $D_f$ and $D_n$, while
$\Delta$ is the superconducting pairing potential in the S layer.
The $(-1)^k$ factor before $I$ accounts for the alternating
polarization of the FM layers. The Usadel equations are supplemented
by the self-consistency equation
\begin{equation}
\Delta \ln \bigg{(}\frac{T_0}{T}\bigg{)}=\pi k_B T \sum_{\omega_l}
\bigg{(}\frac{\Delta}{\hbar |\omega_l|}-F_s\bigg{)},
\label{selfcons1}
\end{equation}
where the summation goes over the Matsubara frequencies $\omega_l =
(2l + 1)\pi k_B T/\hbar$ with all integers $l$. $T_0$ is the $T_c$
of bare S of thickness $d_{s}$, i.e. see Fig.\ref{TcCalibration}.
The anomalous function $F$ obeys the boundary conditions
\begin{equation}
\sigma_a \frac{d F_a(x_{a,b})}{d x}=\sigma_b \frac{d F_b(x_{a,b})}{d
x}=\frac{F_b(x_{a,b})-F_a(x_{a,b})}{R_{a,b}} \label{boundary1}
\end{equation}
at the points $x = x_{a,b}$ separating any $a$ and $b$ layers; $a$
and $b$ can be either of the S, FM or NM layers and in each case we
use the appropriate normal-state conductivities ($\sigma_a$ and
$\sigma_b$) as well as the appropriate interfacial resistance per
unit area $R_{a,b}$ between $a$ and $b$ layers. Since all the FM and
NM layers are identical, the three conductivities are $\sigma_s$,
$\sigma_f$ and $\sigma_n$ and the two interfacial resistances are
$R_{s,f}$ and $R_{f,n}$.

If we introduce the formal vector $\mathbf{\Phi}_b=(F_b, dF_b/dx)$,
we can turn Eq. (\ref{usadelf}) and (\ref{usadeln}) into vector
equations for $\mathbf{\Phi}_b$ ($b$ refers to an arbitrary layer
again), which take the common form
\begin{equation}
\frac{d}{dx}\mathbf{\Phi}_b = \mathbf{M}_b \mathbf{\Phi}_b \quad
\textrm{with} \quad \mathbf{M}_b= \left( \begin{array}{ccc} 0 & 1 \\
k_b^2 & 0
\end{array} \right), \label{matrixeq}
\end{equation}
where the values of $k_b^2$ are
\begin{equation}
k_{fk}^2=\frac{2\hbar |\omega_l|+2i(-1)^k I
\mathrm{sgn}(\omega_l)}{\hbar D_f} \,\,\, \textrm{and}\,\,\,
k_{nk}^2=\frac{2|\omega_l|}{D_n} \label{kb2}
\end{equation}
in the $k$th FM and the $k$th NM layers, respectively. Eq.
(\ref{matrixeq}) is a simple linear differential equation. If
$\mathbf{\Phi}_b$ is known at the left side ($x=x_{a,b}$) of a
layer, its value at the right side ($x=x_{a,b}+d_b$) can be
expressed in terms of matrix exponentials
\begin{equation}
\mathbf{\Phi}_b (x_{a,b}+d_b) = \mathbf{A}_b \mathbf{\Phi}_b
(x_{a,b}) \,\,\,\, \textrm{with} \,\,\,\, \mathbf{A}_b =
\exp(\mathbf{M}_b d_b). \label{matrixexp}
\end{equation}
The boundary conditions (Eq. \ref{boundary1}) between any two layers
$a$ and $b$ can be written in the vector form as
\begin{equation}
\mathbf{\Phi}_b (x_{a,b}) = \mathbf{B}_{b,a} \mathbf{\Phi}_a
(x_{a,b}) \quad \textrm{with} \quad \mathbf{B}_{b,a} = \left(
\begin{array}{ccc} 1 & R_{a,b} \sigma_a \\ 0 & \sigma_a / \sigma_b
\end{array} \right).
\label{boundary3}
\end{equation}
The derivative of the anomalous function $F$ vanishes at the left
end of the multilayer, therefore $\mathbf{\Phi}_{fN}(-d_f
N-d_n(N-1)) =(C,0)$, where $C$ is a complex number. The formal
vector $\mathbf{\Phi}$ at the left side of the S layer is now
expressed in the function of $C$ by systematically going through the
layers and using the appropriate matrices appearing in Eqs.
(\ref{matrixexp}) and (\ref{boundary3}):
\begin{equation}
\mathbf{\Phi}_s (0) = \mathbf{L}(\omega_l) \left( \begin{array}{ccc}
C \\ 0 \end{array} \right), \quad \textrm{where} \label{phi1}
\end{equation}
\begin{equation}
\mathbf{L}(\omega_l) = \mathbf{B}_{s,f1}
\prod_{k=1}^{N-1}[\mathbf{A}_{fk} \mathbf{B}_{fk,nk} \mathbf{A}_{nk}
\mathbf{B}_{nk,f(k+1)}]\mathbf{A}_{fN} \label{phi2}
\end{equation}
is a complex matrix only dependent on material parameters and the
Matsubara frequency $\omega_l$, which depends on $T$.

To obtain $T_c$, we apply the multimode method developed by Fominov
\emph{et al.} \cite{FominovPRB660145072002}. If we write down the
components of the vector Eq. (\ref{phi1}) and eliminate $C$, we
obtain
\begin{equation}
\frac{dF_s(0)}{dx} = Q(\omega_l) F_s(0) \quad \textrm{with} \quad
Q(\omega_l) = \frac{L_{21}(\omega_l)}{L_{11}(\omega_l)},
\label{qdef}
\end{equation}
which is analogous to Eq. (8) in \cite{FominovPRB660145072002}. From
here, the problem is reduced to finding the highest $T$ for which
the determinant of a matrix $\mathbf{K}$ is zero. If we include $M$
modes ($m=1,2,...M$) in addition to the single-mode method and take
the first $M+1$ Matsubara frequencies ($l=0,1,...M$), the elements
of $\mathbf{K}$ are
\begin{equation}
K_{l0} = \frac{R(\omega_l)\cos(k_0 d_s)-k_0 \sin(k_0 d_s)}{\hbar
(\omega_l + k_0^2 D_s /2)} \quad \textrm{and} \label{k0def}
\end{equation}
\begin{equation}
K_{lm} = \frac{R(\omega_l)\cosh(k_m d_s)+k_m \sinh(k_m d_s)}{\hbar
(\omega_l - k_m^2 D_s /2)}, \label{kmdef}
\end{equation}
where $R(\omega_l)$ is given by
\begin{equation}
R(\omega_l) = \mathrm{Re}(Q(\omega_l)) +
\frac{[\mathrm{Im}(Q(\omega_l))]^2}{\mathrm{Re}(Q(\omega_l)) + k_s
\tanh(k_s d_s)} \label{rdef}
\end{equation}
in the function of  $Q(\omega_l)$ and $k_s = \sqrt{2\omega_l/D_s}$.
The quantities $k_0$ and $k_m$ ($m=1,2,...M$) are the smallest
positive roots of
\begin{equation}
\ln \bigg{(}\frac{T_0}{T}\bigg{)} = \psi \bigg{(}\frac{1}{2} +
\frac{\hbar k_0^2 D_s}{4\pi k_B T}\bigg{)}-\psi
\bigg{(}\frac{1}{2}\bigg{)} \quad \textrm{and} \label{selfcons20}
\end{equation}
\begin{equation}
\ln \bigg{(}\frac{T_0}{T}\bigg{)} = \psi \bigg{(}\frac{1}{2} -
\frac{\hbar k_m^2 D_s}{4\pi k_B T}\bigg{)}-\psi
\bigg{(}\frac{1}{2}\bigg{)}, \label{selfcons2m}
\end{equation}
which are obtained from the self-consistency Eq. (\ref{selfcons1})
and contain the digamma function $\psi$.

Using this method, we determine $T_c$ numerically; the calculation
is repeated with various values of $T$, for $0<T<T_0$. By obtaining
the roots $k_0$ and $k_m$ of Eqs. (\ref{selfcons20}) and
(\ref{selfcons2m}), and evaluating $R(\omega_l)$ through Eq.
(\ref{rdef}), the matrix $\mathbf{K}$ can be found for all $T$. The
largest value of $T$ for which $\det(\mathbf{K})=0$ corresponds to
$T_c$. The multimode method with $M \rightarrow \infty$ is exact but
in most cases the inclusion of $M=$ 8 modes suffices.

To apply this numerical method with the minimum number of adjustable
parameters, we have measured $\sigma$ for Nb, Fe and Cr thin films
for $T<$10 K to be 1.9$\times$10$^6$ $(\Omega m)^{-1}$,
6.6$\times$10$^6$ $(\Omega m)^{-1}$, and 2.2$\times$10$^6$ $(\Omega
m)^{-1}$, respectively. From these values, we estimate the electron
mean free paths of Fe and Cr via $\ell= \sigma m v_F.(ne^2)^{-1}$
with $v_F$ the Fermi velocity; $v_F$ for Fe \cite{Covo} and Nb
\cite{Finnemore} is taken from literature, while for Cr we assume a
similar $v_F$ to Fe. The density number of electrons is
$n=(8\pi/3).(mv_F/h)^3$ with $m$ the electron mass, giving an $\ell$
of 2.7 nm and 0.9 nm for Fe and Cr, respectively. For Nb, we
determine $\ell$ by choosing a value that gives the best fit (2.2
nm). With $\ell$ known, we calculate $D$ for Nb, Fe, and Cr via
$D=\ell v_F/3$, giving 2.2$\times$10$^{-4}$ $m^2 s^{-1}$ for Nb,
18$\times$10$^{-4}$ $m^2 s^{-1}$ for Fe, and 6.0$\times$10$^{-4}$
$m^2 s^{-1}$ for Cr. Finally, we estimate the coherence lengths of
Nb and Cr with $\xi_{s,n}=\sqrt{\hbar D_{s,n}/2\pi k_B T_0}$, giving
5.8 nm for Nb (similar to \cite{CoherenceLengNb}) and 9.6 nm for Cr,
while for Fe $\xi_f =\sqrt{\hbar D_f /I}$, giving 3.7 nm (similar to
\cite{Robinson2007}) assuming $I\sim$ 1000 K \cite{Kittel}. For both
Nb and Fe, $\xi>\ell$, which is indicative of dirty limit behavior
and, therefore, justifies our use of the linearized Usadel
equations. For the interfacial resistances we take $R_{Fe,Nb}\sim$
10$^{-15}$ $\Omega m^2$ and $R_{Fe,Cr}\sim$ 10$^{-17}$ $\Omega m^2$
(the quality of the fit does not depend strongly on these
parameters). The important material parameters used/calculated here
are listed in Table \ref{table}. With these values, the numerical
model agrees well with the experimental data; see Fig. \ref{TcvN}.

We have generalized the model to consider the behavior when the Fe
layers are non-collinear; we use the linearized Usadel equations
containing both the singlet and triplet components of the anomalous
Green function \cite{Houzet2007}. We find that with a small change
in $2\theta$, the angle between the polarizations of adjacent Fe
layers, the $T_c$ is more strongly reduced and the parity dependent
oscillations fade away faster; see Fig. \ref{TcvN}(a,i). In the
limiting case of $2\theta=\pi$ we recover the SAF behavior. An
applied magnetic field will produce a non-collinear configuration;
however, in our samples this could not be achieved without directly
suppressing the Nb $T_c$. With improved control of $J$ the field
required to reorient the layers could be substantially reduced.

In conjunction with the experimental results, this demonstrates that
a large change in $T_c$ can be obtained by switching the SAF between
P and AP configurations. In the particular case of $N=$ 2, which
corresponds to the Oh \emph{et al.} \cite{Sangjun} spin valve, the
change in $T_c$ from AP to P is $>$ 1 K, which is a many times
higher than the changes experimentally observed in the analogous
superconductor PSV structures \cite{PSV}.

This Letter has shown that the parity of AFs with perfect order have
a profound effect on the proximity effect as predicted by Andersen
\emph{et al.} and that the results can be well described by the
adaptation of the linearized Usadel equations to this new situation.
However, there are important aspects of the results which are not
explained on this basis. Firstly, there appears to be a longer
period oscillation in $T_c$, which is visible in both data sets as
an upturn in the trend in $T_c$ for $N>$ 5. Perhaps, more
significantly, there appears to be a parity-dependence of the
resistive transition width ($\Delta T_c$) as shown in Fig.
\ref{TcvN}(c), which suggests that the nature of the superconducting
transition is being affected.

\begin{table}[h]
\begin{tabular*}{0.48\textwidth}{@{\extracolsep{\fill}}l c c c c c c } \hline \hline
 &   $\xi$    & $v_F$      &  $I$       &     $\sigma$        &                                     D                &   $\ell$  \\
       &     nm     & $\times$10$^{6}$ ms$^{-1}$  &   K    & $\times$10$^{6}$ $(\Omega m)^{-1}$   &                     $\times 10^{-4}$ m$^2$s$^{-1}$    &    nm    \\
 \hline
Nb    & \emph{\textbf{5.8}}                              & 0.3               & $\ldots$        &  1.9  &          \emph{\textbf{2.2}} & 2.2 \\
Fe    & \emph{\textbf{3.7}}                              & 2.0               & 1000            & 6.6   &          \emph{\textbf{18}}  & \emph{\textbf{2.7}} \\
Cr    & \emph{\textbf{9.6}}                              & 2.0               & $\ldots$        & 2.2   &          \emph{\textbf{6.0}} & \emph{\textbf{0.9}} \\
 \hline  \hline

\end{tabular*}
\caption{Important parameters used/\emph{\textbf{calculated}} in
this letter. \label{table}}
\end{table}

This work was supported by EPSRC UK. We thank Professor James Annett
for theoretical discussions.\\



\begin{references}

\bibitem{Review} For reviews: A. I. Buzdin, Rev. Mod. Phys. \textbf{77}, 935
(2005); F. S. Bergeret, A. F. Volkov, and K. B. Efetov, Rev. Mod.
Phys. \textbf{77}, 1321 (2005).

\bibitem{Jiang1995} J. S. Jiang \emph{et al.}, Phys. Rev. Lett. \textbf{74}, 314 (1995).

\bibitem{Andersen} Brian M. Andersen \emph{et al.}, Phys. Rev. Lett. \textbf{96}, 117005 (2006).

\bibitem{SAFS} C. Bell \emph{et al.}, Phys. Rev. B \textbf{68}, 144517 (2003); Y. Cheng and
M.B. Stearns, J. Appl. Phys. \textbf{67}, 5038 (1990).

\bibitem{ParkinPRL6423041990} S. S. P. Parkin \emph{et al.}, Phys. Rev. Lett. \textbf{64}, 2304 (1990).

\bibitem{FeMulti} P. Koorevaar \emph{et al.}, Phys. Rev. B
\textbf{49}, 441 (1994); T. M\"{u}hge \emph{et al.}, Phys. Rev.
Lett. \textbf{77}, 1857 (1996); M. V\'{e}lez \emph{et al.}, Phys.
Rev. B \textbf{59}, 14659 (1999).

\bibitem{PSVTHEORY} P. G. de Gennes, Phys. Lett. \textbf{23},10 (1966); L. R. Tagirov, Phys. Rev. Lett. \textbf{83}, 2058 (1999); A. I. Buzdin \emph{et al.}, Europhys. Lett.
\textbf{48}, 686 (1999).

\bibitem{PSV} Ion C. Moraru \emph{et al.}, Phys. Rev. Lett. \textbf{96}, 037004 (2006); J. Y. Gu \emph{et al.}, Phys. Rev. Lett. \textbf{89}, 267001 (2002); A. Potenza
and C. H. Marrows, Phys. Rev. B \textbf{71}, 180503(R) (2005).

\bibitem{Sangjun} Sangjun Oh \emph{et al.}, Appl. Phys. Lett. \textbf{71}, 2376 (1997).

\bibitem{Robinson2007} J. W. A. Robinson \emph{et al.}, Phys. Rev. B \textbf{76}, 094522 (2007).

\bibitem{FominovPRB660145072002} Ya. V. Fominov \emph{et al.}, Phys. Rev. B \textbf{66}, 014507 (2002).

\bibitem{Covo} M. K. Covo \emph{et al.}, Phys. Rev. ST Accel. Beams \textbf{9}, 063201 (2006).

\bibitem{Finnemore} D. K. Finnemore \emph{et al.}, Phys. Rev. \textbf{149}, 231 - 243 (1966).

\bibitem{CoherenceLengNb} A. S. Sidorenko \emph{et al.}, Ann. Phys. (Berlin) \textbf{12}, 37 (2003).

\bibitem{Kittel} C. Kittel, \emph{Introduction to Solid State Physics} (John Wiley \& Sons, Inc., New York, 1956).

\bibitem{Houzet2007} M. Houzet and A. I. Buzdin, Phys. Rev. B \textbf{76}, 060504(R) (2007).


\end{references}
\end{document}